\documentclass[times]{aa}
\usepackage{psfig}
\usepackage{graphicx}
\usepackage{natbib}

\bibpunct{(}{)}{;}{a}{}{,}

\begin{document}

\def\eps{\varepsilon}
\def\aap{A\&A}
\def\apj{ApJ}
\def\apjs{ApJS}
\def\apjl{ApJL}
\def\mnras{MNRAS}
\def\aj{AJ}
\def\nat{Nature}
\def\aaps{A\&A Supp.}
\def\prd{Phys. Rev. D}
\def\prl{Phys. Rev. Lett.}
\def\araa{ARA\&A}       

\def\me{m_\e}
\def\lesssim{\mathrel{\hbox{\rlap{\hbox{\lower4pt\hbox{$\sim$}}}\hbox{$<$}}}}
\def\gtrsim{\mathrel{\hbox{\rlap{\hbox{\lower4pt\hbox{$\sim$}}}\hbox{$>$}}}}

\def\vr{\vec{r}}
\def\vrp{\vec{r}_\perp}
\def\vp{\vec{v}_\perp}
\def\ii{\hat{\i}}
\def\btheta{{\boldmath{$\hat{\theta}$}}}
\def\vx{{\bf x}}
\def\vxp{{\bf x'}}
\def\vy{{\bf y}}

\def\del#1{{}}

\def\C#1{#1}

\title{Measuring Dark Matter Flows in Merging Clusters of Galaxies}
\titlerunning{Measuring Dark Matter Flows in Merging Clusters of Galaxies}
\author{J.~A. Rubi\~no-Mart\'{\i}n \and C. Hern\'andez-Monteagudo \and T.~A. En{\ss}lin}
\authorrunning{J. A. Rubi\~no-Mart\'{\i}n}
\institute{Max-Planck-Institut f\"{u}r Astrophysik,
Karl-Schwarzschild-Str.1, Postfach 1317, 85741 Garching, 
Germany} 
\date{Received date..............; accepted date................}

\abstract{
The Rees-Sciama effect produced in mergers of galaxy clusters 
is discussed, and an analytical approximation
to compute this effect from numerical simulations is given.
Using this approximation and a novel toy model describing the physics
of the merger, we characterize the spatial properties and symmetries
of the Rees-Sciama signal. 
Based on these properties, we propose a method 
to extract the physical parameters of the merger, which
relies on the computation of the quadrupole moment 
of the observed brightness distribution on the sky. 
The relationships between the quadrupole coefficients and
the physical parameters of the merger (physical separation, 
projection angle on the sky and angular momentum) are discussed.
Finally, we propose a method to co-add coherently the RS signals 
from a sample of cluster mergers, in order to achieve an statistical
detection of the effect for those cases where individual signals  
are masked by the kinetic SZ effect, the primordial
CMB components, and by observational noise.
 } 
\maketitle

\section{Introduction\label{sec:intro}}

The integrated Sachs-Wolfe \citep[ISW;][]{1967ApJ...147...73S,
1994PhRvD..50..627H} effect produces CMB temperature fluctuations due
to the accumulation of red- or blue-shift of photons travelling in
time dependent gravitational potentials. Several recent
studies
\citep{2003astro.ph..5001B,2003astro.ph..5097N,2003astro.ph..5468F} 
have claimed evidence for its detection 
in the WMAP data \citep{bennett03}.
The non-linear contribution
to the ISW effect, in which the density contrast producing the
gravitational potential is in its non-linear regime, 
is usually called the Rees-Sciama 
\citep[RS;][]{rees_sciama68, seljak96} effect.

Here we discuss the RS effect in the extremely non-linear regime of
present day galaxy cluster mergers. The regime of moving single galaxy
clusters was already examined by \citet{birkinshaw83}, 
\citet{1998A&A...334..409A},
\citet{2000ApJ...537..542M, 2003ApJ...586..731M}, and
\citet{2002PhRvD..65h3518C}.  However, galaxy clusters reach the
largest velocities, and therefore the strongest RS signal, during
mergers in which two or more clusters are invoked.  Due to the slow
centre-of-mass velocity of the merging system, the RS signals of the
merging subclusters can partly cancel or increase each other,
depending on the merger geometry and the location of the
line-of-sight (LOS).

Furthermore, the spectral signature of the RS effect is a temperature
change of the measured CMB photons, and therefore indistinguishable
form the kinematic Sunyaev-Zeldovich 
\citep[kSZ;][]{1972ComAp...4..173S,1980ARA&A..18..537S}
effect spectrally. However,
both effects have distinct morphologies, which in principle allow to
discriminate them. 

The focus of this work is to examine methods to extract the RS
signature from merging pairs of galaxy clusters.
The layout of the paper is as follows. In Sec. 2 we provide 
an approximation to compute the RS effect either from 
theoretical cluster models or numerical simulations. 
In addition, a brief description of the kSZ and the thermal
Sunyaev-Zeldovich (tSZ) effects is also given, emphasizing their
phenomenological differences with the RS effect.
In Sec. 3 we present and use an analytic toy model of a merger of two 
galaxy clusters to characterize the typical
morphologies of these effects (RS, kSZ and tSZ). 
Based on its spatial properties, we present in Sec. 4 a method
to extract both the RS signal and the physical parameters describing
the system from a given merger.
A method to stack the signal from a sample of mergers is also proposed.
Conclusions are presented in Sec. 5.

\section{Formalism}
\subsection{Rees Sciama effect}

The CMB temperature change in a LOS direction $\vec{\hat{n}}$ caused
by the RS effect can be written as \citep{2002PhRvD..65h3518C}
\begin{equation}
\label{eq:dT1}
\delta T_{\rm RS}(\vec{\hat{n}}) = - \frac{2}{c^3} \, \int \! dr\,
\dot{\Phi}(\vec{\hat{n}}\,r, r/c)\,T(\vec{\hat{n}}) ,
\end{equation}
where $\dot{\Phi}(\vr,t)$ is the time derivative of the
gravitational potential and $r/c$ the look-back time along the LOS.

In the following, we are dealing with well separated structures,
galaxy clusters, which do not change significantly during the time
of a photon passage. Thus, we explicitly 
neglect the intrinsic variation of the cluster potential while
the photon is passing through it. 
This allows an approximative treatment.

In Newtonian approximation, the gravitational potential depends on the
total mass density $\varrho(\vr,t)$ via
\begin{equation}
{\Phi}(\vr,t) = -\,G\, \int\! d^3 r' \,
\frac{\varrho(\vr',t)}{|\vr-\vr'|}.
\end{equation}
The time derivative of the gravitational potential therefore
follows the change of the matter density, leading to
\begin{equation}
\label{eq:phidot}
\dot{\Phi}(\vr,t) = -\,G\,\int\! d^3r' \, \varrho(\vr',t)\,
 \vec{v}(\vr',t) \vec{\cdot}
\frac{\vr-\vr'}{|\vr-\vr'|^3}
\end{equation}
by using the continuity Eq. $\dot{\varrho} +\vec{\nabla}
(\vec{v}\,{\varrho}) = 0$, then the Gauss' theorem, and neglecting
boundary terms in the integration. Here, $\vec{v}$ is the mean velocity
of matter, which is the density weighted mean of all matter species'
velocities $\vec{v} = \sum_i \,\vec{v}_{i} \,\varrho_{i}/\varrho$, where 
the $i$ runs over all species like dark matter, gas, and galaxies, and
$\varrho =  \sum_i \,\varrho_{i}$.

By inserting Eq. \ref{eq:phidot} into Eq. \ref{eq:dT1} we approximate
all LOS to be parallel to the $z$-direction, we ignore the explicit
dependency of $\varrho$ and $v$ on time, and integrate out the LOS
integral, yielding
\begin{equation}
\label{eq:dTRS2}
\frac{\delta T_{\rm RS}}{T} (\vec{r_\perp}) = -\frac{4\,G}{c^3}\, 
\int d^3 r' \varrho(\vr')\, \vec{v}(\vr') \vec{\cdot} \frac{\vrp -
\vrp'}{|\vrp - \vrp'|^2},
\end{equation}
where vectors are split into a LOS-parallel and perpendicular part
according to $\vr = (\vrp, z)$, and where the Z-axis is taken pointing 
towards the observer. This approximative
description of the RS effect ignores small corrections due to
non-parallel LOSs, gravitational lensing, the actual change of the
matter distribution during the photon passage, and the finite velocity
of gravitational interaction.

Nevertheless, its accuracy is sufficient to study the general
difficulties in RS signal extraction from merging clusters of
galaxies. Its simplicity allows easy inclusion into any code which
calculates the kSZ effect from numerical simulations.

One can introduce the LOS projected momentum 
\begin{equation}
\tilde{\vec{p}}(\vrp) = \int\! dz \, \varrho(\vrp,z)\, \vec{v}(\vrp, z)
\end{equation}
in order to write
\begin{equation}
\label{eq:dTRS3}
\frac{\delta T_{\rm RS}}{T} (\vec{r_\perp}) = -\frac{4\,G}{c^3}\, 
\int d^2 r'_\perp \, \tilde{\vec{p}}_\perp(\vrp') \vec{\cdot} \frac{\vrp -
\vrp'}{|\vrp - \vrp'|^2}.
\end{equation}
From this, one easily sees that the RS effect measures the convergence
of the projected perpendicular momentum.

\subsection{Moving cluster of galaxies effect}

Eq. (\ref{eq:dTRS3}) can also be reproduced using a different approach. The
moving cluster of galaxies (MCG) effect \citep{birkinshaw83} was
originally derived as a gravitational lensing effect. However, and as
it was pointed out by \cite{2000ApJ...537..542M}, this effect can 
be seen as an special type of RS effect, 
which is not caused by intrinsic variations
of the gravitational field, but due to the movement of the cluster
with respect to the rest frame of the CMB.  Given that we have used
exactly this assumption when deriving equation (\ref{eq:dTRS2}), then
our expression is completely explained as a gravitational lensing
effect.

Following \cite{birkinshaw83} and \cite{2000ApJ...537..542M}, the MCG
effect for a cluster moving with velocity $\vec{v}$, for small values
of the lensing deflection angle, can be written as
\begin{equation}
\frac{\delta T_{MCG}}{T}(\hat{n}) = - \frac{v}{c} \sin \alpha \,
 \vec{\hat{u}}_{v \perp} \cdot \vec{\delta}(\hat{n})
\label{eq:mcg1}
\end{equation}
where $\alpha$ is the angle between the cluster's peculiar velocity
vector and its angular position vector, $\vec{\hat{u}}_{v \perp}$ is
an unit vector parallel to the projection of the velocity in the lens
plane ($\vp = v \sin \alpha \, \vec{\hat{u}}_{v \perp}$), and
$\vec{\delta}(\hat{n})$ is the vector describing the deflection angle.
If we now use the Eq. for a geometrically-thin lens from
\cite{schneider_book}, 
\begin{equation}
\vec{\delta} (\vec{r_\perp}) = \frac{4\,G}{c^2}\, \int d^2 r'
\Sigma(\vrp')\, \frac{\vrp - \vrp'}{|\vrp - \vrp'|^2},
\label{eq:mcg2}
\end{equation}
where $\Sigma$ is the surface mass density, and we insert it in
Eq. \ref{eq:mcg1}, we recover Eq. \ref{eq:dTRS3} in its limit of a
constant velocity field.  Since the MCG effect is linear, one can
build a superposition of differently moving matter distributions, and
therefore fully recover the more general Eq.  \ref{eq:dTRS3}.  It
should be noted that Eq. \ref{eq:mcg2} requires that the gravitational
fields should be weak (which is true for clusters of galaxies), and
that the matter distribution of the lens must be nearly 
stationary (i.e. $v/c \ll 1$).

\subsection{Kinematic SZ effect}

Since the kSZ effect has the same spectral signature as the RS effect,
it is crucial to include the latter into the considerations. Using similar
planar and time-independent approximations the kSZ effect reads
\begin{equation}
\frac{\delta T_{\rm kSZ}}{T} (\vrp) = \int\! dz \,
\sigma_{\rm T}\, n_{\rm e}(\vrp,z)\, v_{{\rm e},z}(\vrp,z)/c,
\end{equation}
where $n_{\rm e}$, and $\vec{v}_{\rm e}$ are the electron density and
velocity, respectively. 

Since the electron and dark matter velocity fields are not too
different, and the baryonic matter density follows roughly the dark
matter density ($\varrho_{\rm gas}(\vr) \approx f_{\rm b}\,
\varrho(\vr)$, with $f_{\rm b}$ the cosmic baryon mass
fraction), one can state that
\begin{equation}
\label{eq:dTkSZ}
\frac{\delta T_{\rm kSZ}}{T} (\vrp) \approx 
\frac{\sigma_{\rm T}\,f_{\rm b}}{m\,c} \, \tilde{\vec{p}}_{z}(\vrp),
\end{equation}
where $m= m_{\rm p} \,(1+\frac{1}{4} \,X_{\rm He})/(1+\frac{1}{2}
\,X_{\rm He})$ is the gas mass per electron, $X_{\rm He}=0.24$ the
primordial He gas mass fraction, and $m_{\rm p}$ the proton mass.

Comparing Eq. \ref{eq:dTRS3} with Eq. \ref{eq:dTkSZ} one recognises
properties which hopefully allow to separate the RS signal from the
kSZ effect:
\begin{enumerate}
\item The RS effect is only sensitive to perpendicular momentum,
whereas the kSZ effect traces only parallel momentum. Thus, in a
cluster merger which happens mainly within the sky plane the RS effect
is enhanced relatively to the SZ effect.
\item The kSZ effect is restricted to LOSs with baryonic matter along it,
whereas the RS effect reaches further out due to its gravitational
origin. This can be used to enhance the relative signal strength.
\item The RS has a more complex morphology compared to the kSZ since
  for a single moving mass, the RS has a dipole, whereas the kSZ a
  monopole morphology. For a pair of merging clusters with opposite
  momenta of the same strength this leads to a mostly quadrupole
  morphology of the RS effect and a dipole morphology of the kSZ
  effect.
\end{enumerate}

\subsection{Thermal SZ effect}

Because of its different spectral characteristic the tSZ effect can
be separated from the the kSZ and RS effects. However, it has to be
considered simultaneously not only because of possible residuals from
incomplete removing of the tSZ effect from measurements, but also
since the tSZ effect helps to localise the gas distribution and
therefore to identify the regions which are expected to contain the
strongest kSZ contamination. The usual tSZ $y$-parameter is given by
\begin{equation}
y(\vrp) = \int\! dz \, \sigma_{\rm T}\, n_{\rm e}(\vrp,z)\, 
\frac{ kT_{{\rm e}}(\vrp,z)}{m_{\rm e}\,c^2},
\end{equation}
where $T_{\rm e}$ stands for the electron temperature, 
and measures the LOS thermal gas energy content.

\section{Toy model}

\subsection{Justification of approximations}

In order to be able to study the strength of the signal we construct a
simplified toy model of a binary cluster merger. We grossly simplify
the complicated gravitational and hydrodynamical process in order to
have an analytic model. Nevertheless this model should provide us with
maps of the RS, kSZ, and tSZ effects, which are sufficiently realistic
for the purpose to study RS signatures by describing the
typical strength and morphology of the effects.

We build the merger model out of two cluster models, labelled with the
numbers 1 and 2, and each described by a NFW-DM matter profile
\citep{1997ApJ...490..493N} filled
by isothermal hydrostatic gas. The two clusters approach each
other. Each cluster velocity $\vec{v}_i$ ($i\in\{1,2\}$) is calculated
to be the velocity of a free-falling object in the gravitational
potential of the other cluster. The total momentum of the merging
system is assumed to be negligible.

Deformation of the clusters due to mutual tidal forces are ignored for
simplicity, so that the DM and the gas of each cluster are streaming
with the same velocity. Furthermore, hydrodynamical interactions of
the gases of the two clusters are ignored, the gases are treated as
non-interacting fluids. This coarse approximation is justified in so
far that the kSZ and the RS effects both are only sensitive to
momentum, and only in projection, but not to the thermodynamical state
of the gas. The ignorance of shock waves exchanging the momentum of
the gases therefore does not strongly change the signal strength, only
its exact spatial morphology.

In contrast to that, the tSZ effect is very sensitive 
to the thermodynamical state of the
gas and has therefore to be treated with a little bit more care. Again
we do not try to model the exact spatial structure but try to get a
good estimate of the signal strength. For that, we have to estimate the amount
of shocked gas produced by the merger. We do this by measuring for
each volume element the excess kinetic energy of the gas of the two
superposed clusters with respect to the volume's bulk kinetic energy
and assuming this to be dissipated. The total thermal energy density
of a volume element is therefore approximated to be 
\[
  \eps_{\rm th}(\vr) = 
\]
\begin{equation}
 \sum_{i=1}^2 \Bigg( \frac{3}{2}\,d_{\rm e}\, 
n_{{\rm  e},i}(\vr)\, k T_{\rm e,i}(\vr) +  
 \frac{m}{2} n_{{\rm e},i}(\vr) 
\,(\vec{v}_i(\vr)-\vec{v}_{\rm e}(\vr))^2 \Bigg)
\end{equation}
where $d_{\rm e} = 2\,(1+\frac{3}{8} \,X_{\rm He})/(1+\frac{1}{2}
\,X_{\rm He})$ is the number of particles per electron, 
$T_{e,i}(\vr)$ ($i\in\{1,2\}$) is the temperature of the gas in each   
cluster, and
$\vec{v}_{\rm e}= \sum_{i\in\{1,2\}} \,\vec{v}_i \,n_{{\rm e},i} /\sum_i
\,n_{{\rm e},i}$ is the local mean gas velocity. From this, the
electron pressure entering the tSZ effect estimate is easily
obtained:

\begin{eqnarray}
\nonumber
(n_e\,kT_{\rm e}) (\vr) &=& n_{{\rm e},1}(\vr)\,kT_{\rm e,1}(\vr) + n_{{\rm
  e},2}(\vr)\,kT_{\rm e,2}(\vr) + \\
&& \frac{m}{3\,d_{\rm e}} \frac{n_{{\rm
  e},1}(\vr)\,n_{{\rm e},2}(\vr)}{n_{{\rm e},1}(\vr) + n_{{\rm
  e},2}(\vr)} \, (\vec{v}_1-\vec{v}_2)^2
\end{eqnarray}

\subsection{Simulations of the effect}

In this subsection, we present the equations outlined in the previous
one for the simulation of a merger of two galaxy clusters, 
with a null total linear momentum. We assume a NFW dark matter
profile for each cluster,
\begin{equation}
\varrho_{DM} (r) = \frac{\varrho_0}{(r/r_s)(1+r/r_s)^2}.
\label{ec:nfw}
\end{equation}
The gas is considered isothermal, and distributed in hydrostatic
equilibrium, so we have \citep{1998ApJ...497..555M}
\begin{equation}
\varrho_g(r) = \varrho_{g0} e^{-b} \Bigg( 1 + \frac{r}{r_s} 
\Bigg)^{b r_s /r}
\label{ec:gas}
\end{equation}
where b is a constant for each halo, given by
\begin{equation}
b \equiv \frac{4\pi G \mu m_p \varrho_0 r_s^2}{k_B T_e}
\end{equation}
The value for $\varrho_{g0}$ is 
determined using the following expression for the gas mass fraction
\begin{equation}
\frac{M_{gas}(r_v)}{M_{tot}(r_v)} = \Upsilon \frac{\Omega_b}{\Omega_m}
\label{eq:rho_gas}
\end{equation}
where the masses are evaluated at the virial radius ($r_v$). 
We use here the values
$\Omega_b=0.044$ and $\Omega_m=0.27$ \citep{bennett03}, and
we adopt $\Upsilon = 0.8$ \citep{1998ApJ...497..555M}. 
It should be noted that when deriving Eq. \ref{ec:gas}, and when  
computing the total mass for Eq. \ref{eq:rho_gas}, 
we have neglected the gas contribution to the total density.

For the computation of the RS effect, we use the 2-D projection of the
NFW profile (Eq. \ref{ec:nfw}), which is given by
\[
\Sigma_{NFW}(r_\perp) = \int_{-\infty}^{+\infty} dz \; 
\varrho_{DM} (\sqrt{z^2 + r_\perp^2}) = 
\]
\[
\left \{
\begin{array}{cc}
\frac{2 r_s^3 \varrho_0}{r_s^2-r_\perp ^2} 
\Bigg[ \frac{r_s}{\sqrt{r_s^2-r_\perp ^2}} 
\ln \Bigg( \frac{r_\perp}{r_s-\sqrt{r_s^2-r_\perp ^2}} \Bigg) -1
\Bigg], &  \qquad r_\perp < r_s \\
\frac{2 r_s^3 \varrho_0}{r_\perp ^2-r_s^2} 
\Bigg[ \frac{r_s}{\sqrt{r_\perp ^2-r_s^2}} 
\Bigg( \arcsin(\frac{r_s}{r_\perp}) - \frac{\pi}{2} \Bigg) +1 \Bigg],
& \qquad r_\perp > r_s \\
\frac{2}{3} r_s \varrho_0, & \qquad r_\perp = r_s  \\
\end{array}
\right.
\]

A simulation of two merging clusters is then characterised by the set
of parameters \btheta = $\{ M_1,M_2,d,\ii,\vec{J}\}$, which correspond
to the masses of the two clusters, the true physical separation, the
projection angle of sky ($\ii =90\degr$ is a face-on merger), and the
angular momentum of the system.  Once these parameters are given, we
can derive the remaining quantities describing the system. For
definiteness, we use the following scaling
relations. Both clusters are taken to be isothermal, with electron
temperatures given by \citep{eke96,1999ApJ...515..465A}
\[
T_e = 7.75 \, \Bigg( \frac{M}{10^{15} h^{-1} M_\odot} 
\Bigg)^{2/3} (1+z) ~ {\rm keV}
\]
Virial radii are derived as \citep{2000ApJ...537..542M}
\[ 
r_v = 1.3 h^{-1} \Bigg( \frac{M}{10^{15} h^{-1} M_\odot} 
\Bigg)^{1/3} (1+z)^{-1} ~{\rm Mpc} 
\]
and the $r_s$ parameter is computed as $r_c = 0.22 r_s$ 
\citep{1998ApJ...497..555M}. We have used a constant ratio $r_v/r_c=10$
(self-similar core evolution, \cite{1986MNRAS.222..323K}).
The electron density is derived as 
$n_{e}(r) = \varrho_g(r)(1+X)/(2 m_p)$, where 
we fix the hydrogen abundance $X$ to  the primordial value
($X=0.76$). 

Velocities are estimated following the formalism described in
\cite{2002mpgc.book....1S}, although we modified it to use the
gravitational energy associated to the NFW matter distribution
described by Eq. \ref{ec:nfw}, because it is
more realistic and better behaved at $r=0$.  For this latter profile, the
gravitational potential is found to be
\begin{equation}
\phi(r) = -4\pi G \varrho_0 r_s^3 \frac{\ln(1 + r/r_s)}{r}.
\end{equation}
When estimating from here the gravitational energy associated to
the two clusters, we follow \cite{2002mpgc.book....1S} 
and use the approximation 
$E_{grav} \approx 1/2 (\phi_1 M_2 + \phi_2 M_1)$, where $\phi_i$ is the
gravitational potential of cluster $i$ at the position of the centre
of the other cluster. From this we calculate the relative
velocity as
\begin{equation}
v^2 \approx \frac{2}{m} \Bigg[ E_{grav}(d_0) - E_{grav}(d) \Bigg] 
\Bigg[ 1 - \frac{b^2}{d_0^2} \Bigg]^{-1} 
\end{equation}
where $d$ is the 3-dimensional separation of the clusters, 
$d_0$ is the initial distance from which clusters fall together 
(see Eq. 10 in \citet{2002mpgc.book....1S}), 
$m \equiv M_1 M_2 /(M_1+M_2)$ is the reduced mass
of the system, and $b$ is the impact parameter of the merger.
From here, the individual cluster velocities 
are calculated as $v_1 = M_2/(M_1+M_2)v$ and $v_2 = -M_1/(M_1+M_2)v$,
so that the total momentum is zero and that $v_1 - v_2 = v$.

\subsection{Spatial properties/symmetries}

In this subsection, we use our toy model for two merging clusters to 
present two examples illustrating the spatial properties and
symmetries of the effect. 
For definiteness, we choose the coordinate
system with the Z-plane as the sky plane, 
with positive values pointing towards the observer.
If we define $\alpha_{CM}$ as the angle between the direction 
joining the two centres of mass of the clusters (from $M_1$ to $M_2$)
and the X-axis, then we choose $\alpha_{CM}=0$ (i.e. clusters 
on the X-axis, and $M_1$ on the left of the image). This coordinate system
will be used throughout the paper, unless otherwise stated. 

\begin{figure*}[t]
\centering
\includegraphics[width=5.8cm]{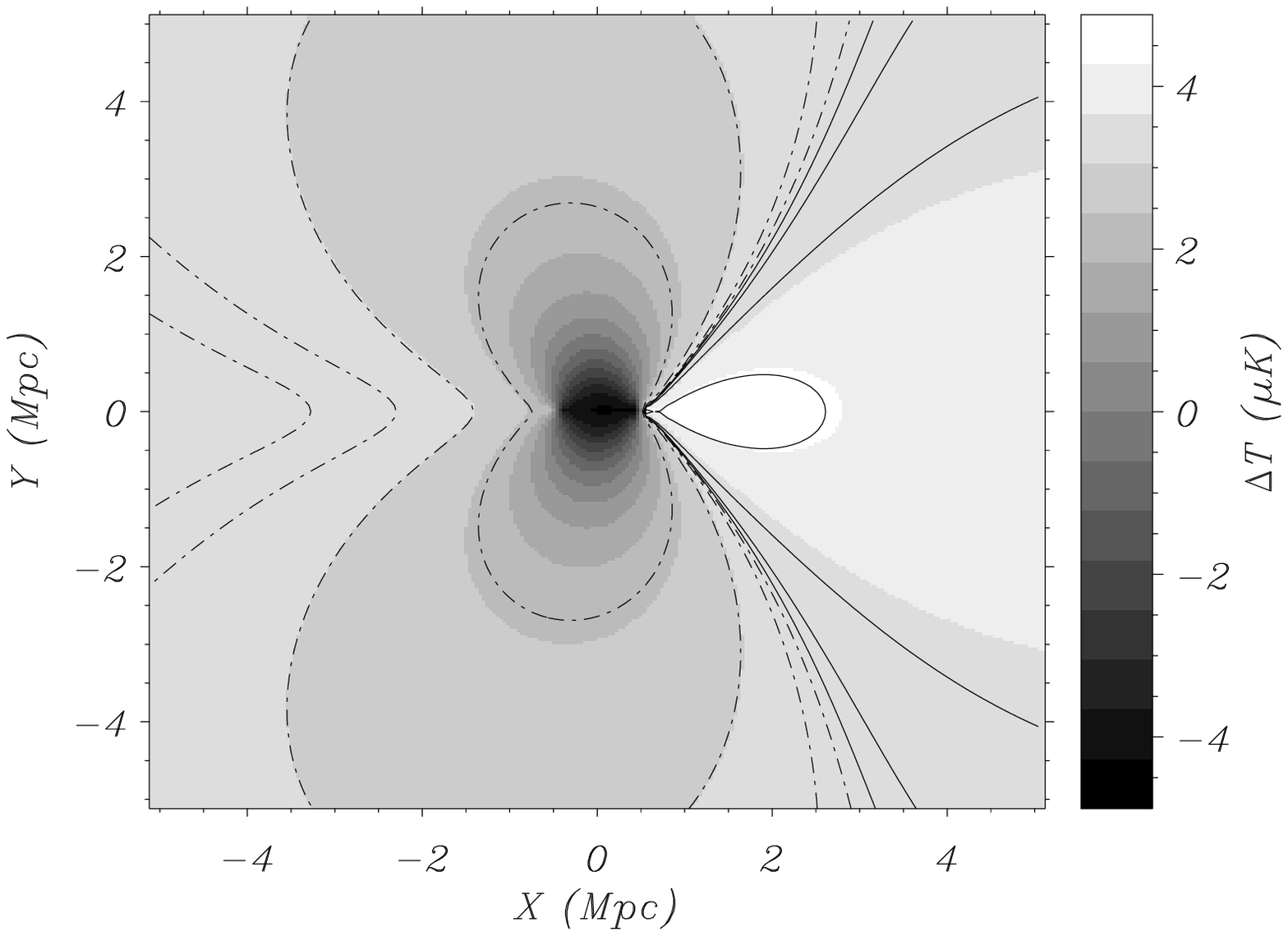}%
\includegraphics[width=5.8cm]{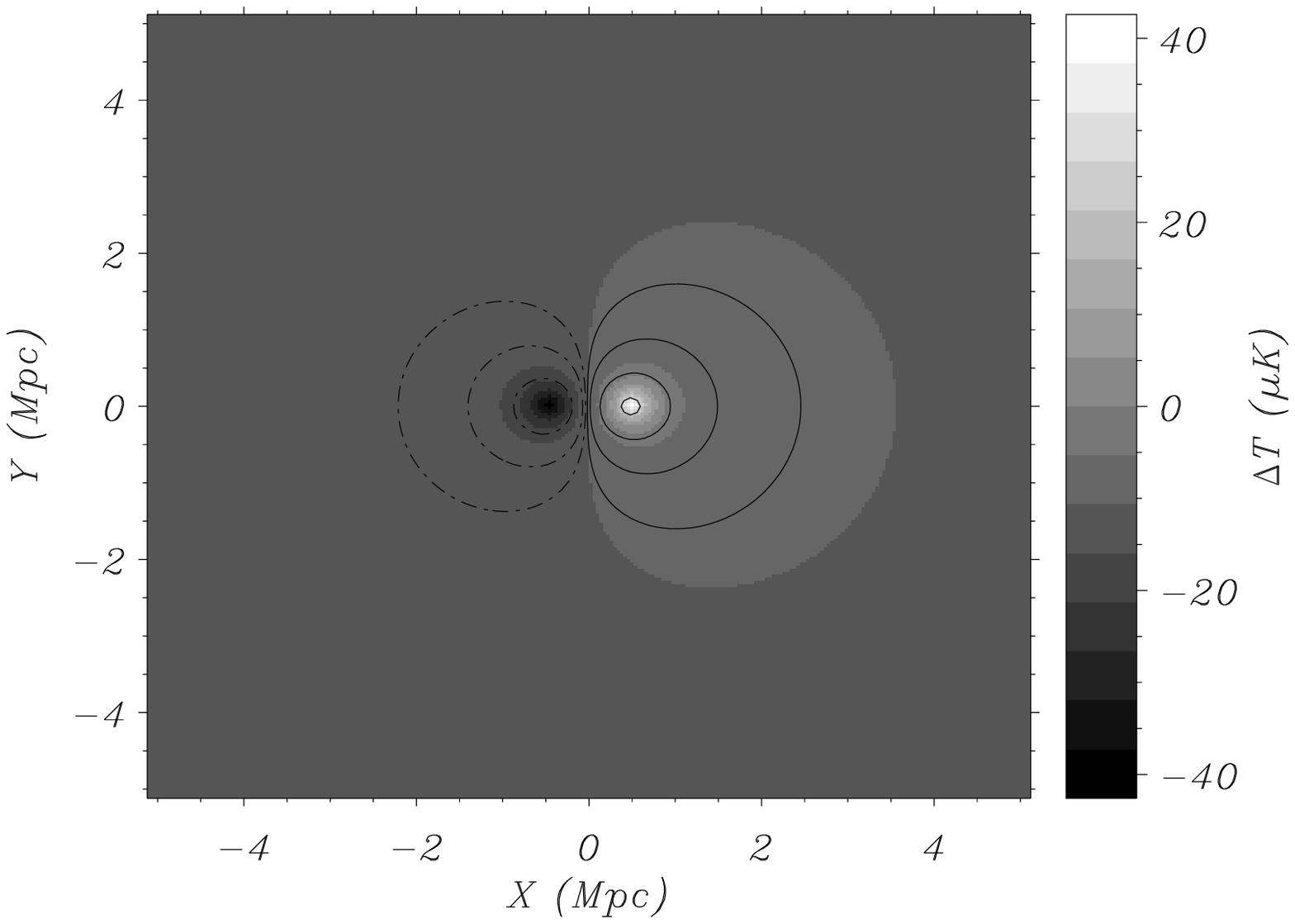}%
\includegraphics[width=5.8cm]{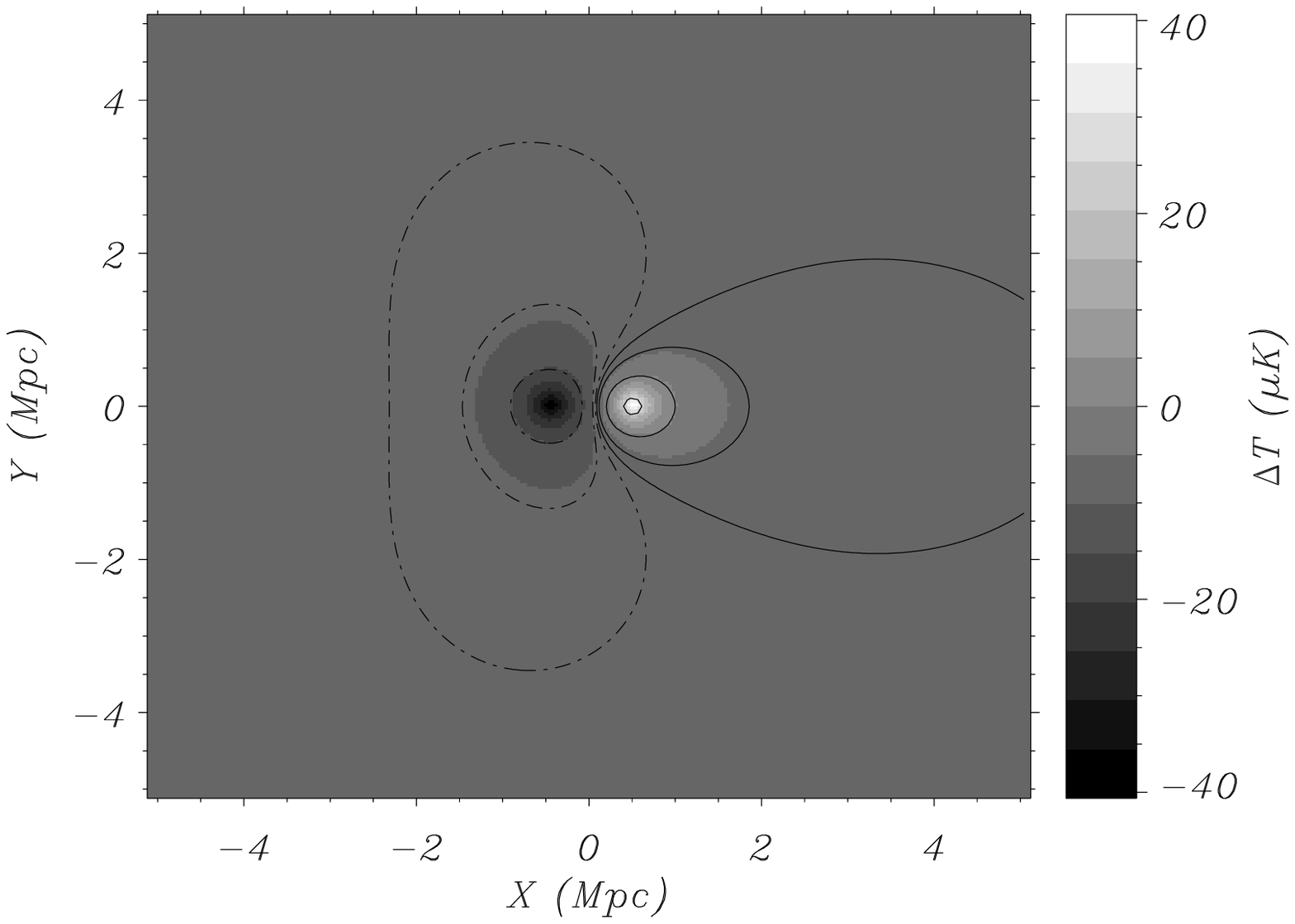}
\caption[]{ RS (left image), kSZ (central image) and the sum of both
effects (right image), for a 
pair of merging clusters with masses $10^{15} M_\odot$ (left cluster) and  
$5\times 10^{14} M_\odot$ (right cluster), with zero
angular momentum, when observed with projection angle $80\degr$, and
a physical separation of 1~Mpc. 
Contour lines in all plots correspond to (-64,-16,-4,-1,1,4,16,64) times the 
maximum of the map (dotted-lines for negative regions and solid lines
for positive ones).
\label{fig:example1} }
\end{figure*}

We first show in Fig. \ref{fig:example1} the result for a
merger with \btheta = $\{ 10^{15} M_\odot, 5\times 10^{14} M_\odot, 
1~{\rm Mpc}, 80\degr, \vec{0}\}$. This is a very optimistic case (in 
terms of the RS signal strength compared to the kSZ effect),
where the  merger is practically face-on.  

As pointed out in Sec. 2.3, the spatial pattern of the RS
effect for a merging of two clusters shows a peculiar
quadrupole structure. This quadrupole structure carries information
about the velocity field of the clusters, and can exhibit, for
example, if both clusters are in a pre- or post- merger state.

This fact is used in the next section to extract the
RS signal in a map where kSZ is also present, given that 
the kSZ has a dipole spatial structure. 
On the other hand, it can also be seen in the figure that the RS effect
has a much more extended pattern than the kSZ, because of its
gravitational nature. 

\begin{figure*}[t]
\centering
\includegraphics[width=6cm]{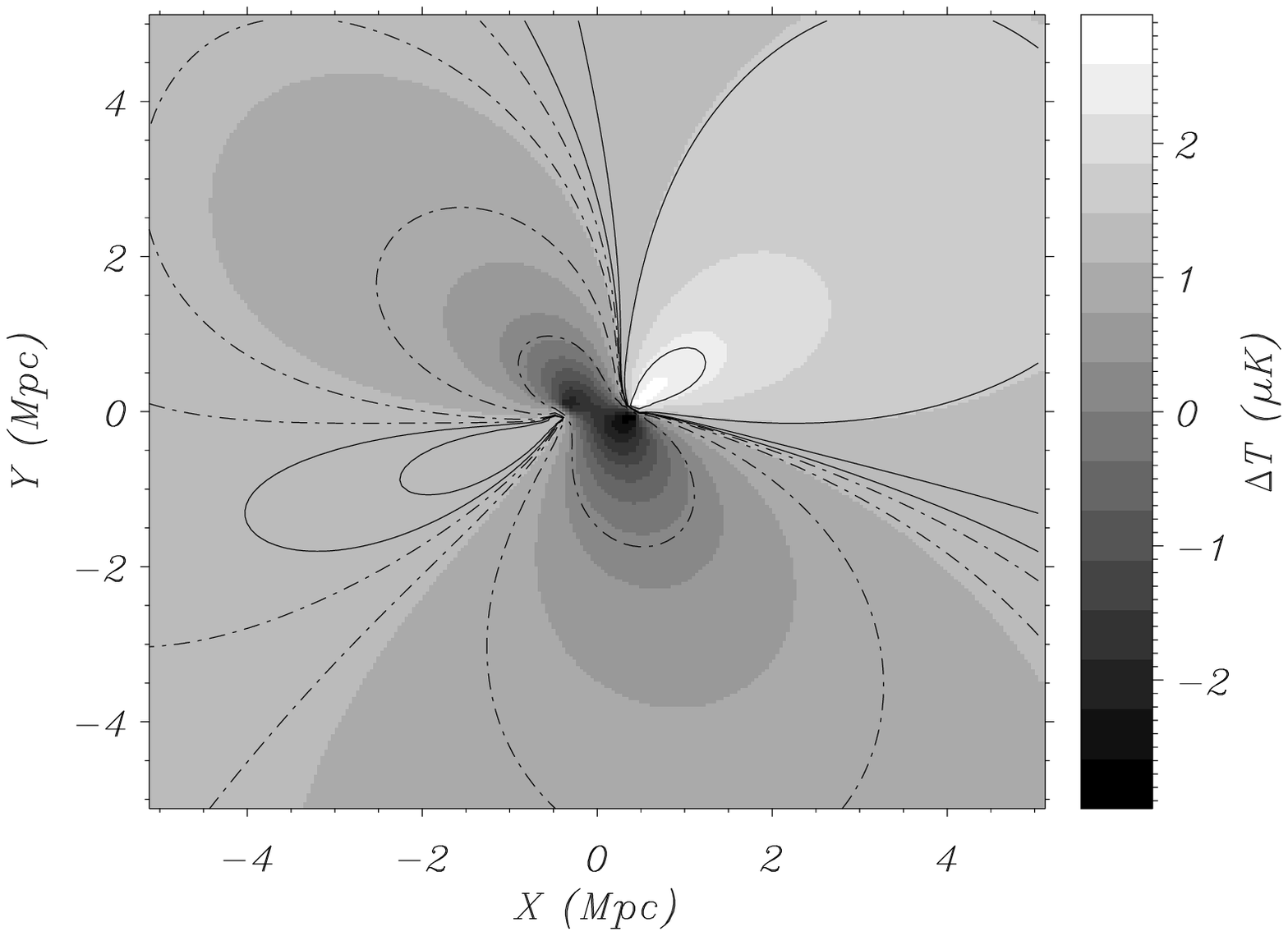}%
\includegraphics[width=6cm]{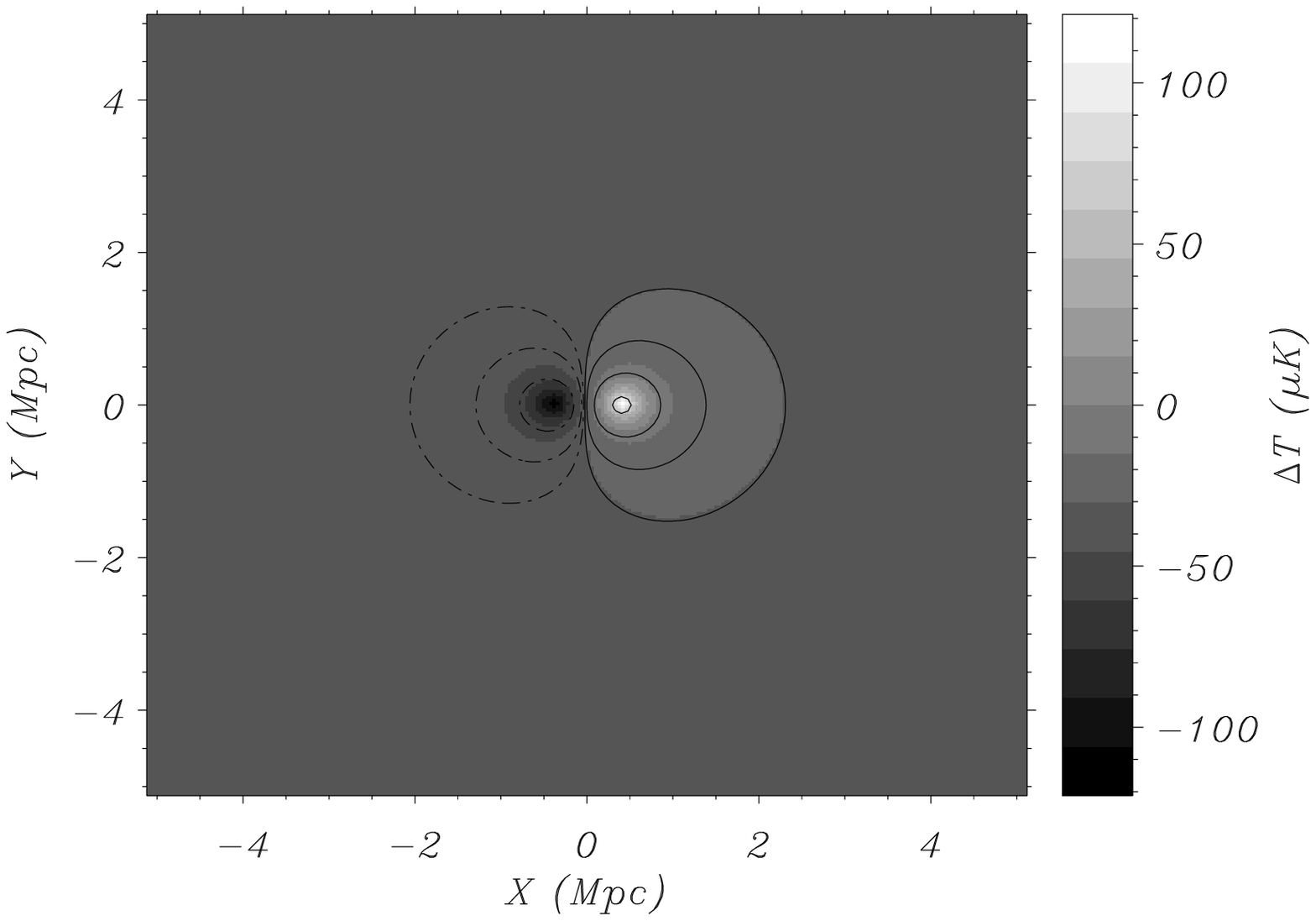}%
\includegraphics[width=6cm]{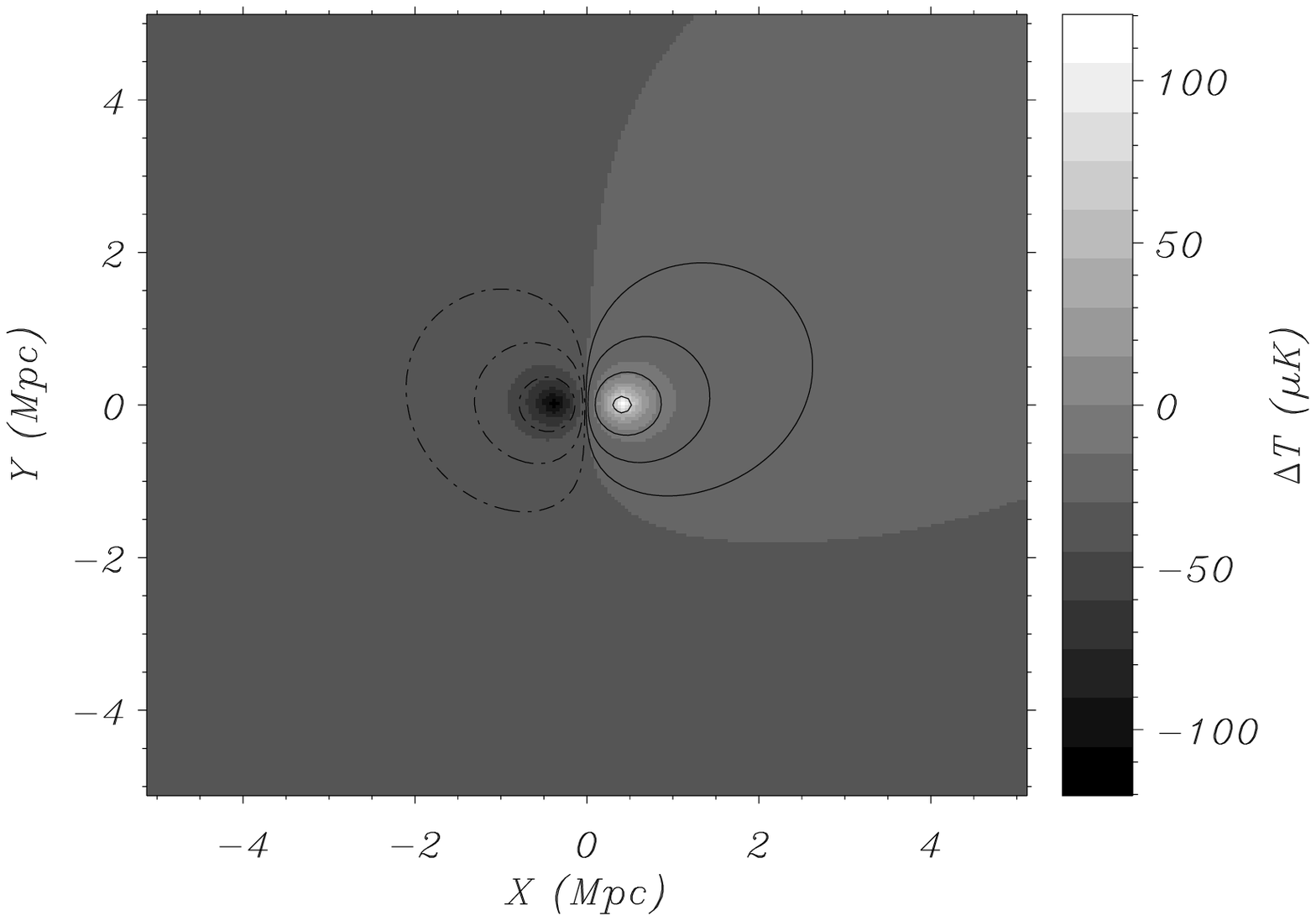}
\caption[]{Same as Fig. \ref{fig:example1}, but considering angular
momentum, and with a larger projection angle. Here, $\ii = 60\degr$,
and the velocity vector is defined by the spherical angles 
$(60\degr,60\degr)$. Contours have the same meaning as in 
Fig.\ref{fig:example1}. 
\label{fig:example2} }
\end{figure*}

As a second example, in Fig. \ref{fig:example2} we consider a
more pessimistic case in which the velocity forms an angle of
$60\degr$ with the line of sight. We incorporate also angular
momentum.  In this case, \btheta = $\{ 10^{15} M_\odot,\, 5\times
10^{14} M_\odot,\, 1~{\rm Mpc},\, 60\degr,\, \vec{J}\}$. Here a
non-zero angular momentum is implicitly defined by assigning 
a velocity direction to the cluster $M_1$ different from the direction
joining the two cluster centres. The direction of $\vec{v}_1$ is
specified by the  
spherical polar angles $(\theta_v,\phi_v) = (60\degr,60\degr)$, where
$\theta_v$ is the usual polar angle from the Z axis, and $\phi_v$
the usual azimuthal angle measured from the X-axis.  In the case 
presented in Fig. \ref{fig:example2}, 
the RS effect is practically hidden by the kSZ, although it can
be marginally seen at large distances.

\section{Signal extraction}

Once we have illustrated the morphology of the RS effect, we will
study how to extract the RS information from observations of
merging clusters. 

\subsection{RS signature detection}

The peculiar structure of the RS effect in merging clusters can be
quantified studying the dipole and quadrupole moments of the observed
intensity map.  
It is worth remarking that the quadrupole moment of the surface
brightness is also used as a tool to extract the shear signal when
studying weak lensing-induced distortion of faint background
galaxies \citep{1995ApJ...449..460K}. 

We propose the following recipe to proceed with
future experiments aiming to measure this effect:
\begin{enumerate}
\item Using a catalogue of cluster mergers, one identifies those
clusters with a suitable projection angle (close to face-on mergers);
\item Observations of the merger in frequencies far from the tSZ
cross-over frequency (217~GHz, or 1.25~mm), or in X-rays,
will show the spatial distribution of the gas. 
One of these images can be used to find the 'centre of
light' point, to which we refer the dipole and quadrupole 
moments\footnote{Note that, in general, the center of X-ray emission,
the center of tSZ-light and the center of mass do not necessarily 
coincide in the same point.}.
In what follows, we will place the origin of our coordinate 
system at this point;
\item Observations at 217~GHz (or performing a spectral component separation 
using observations at several frequencies) will provide us with a  
map where only kSZ and RS effects (as well as the CMB fluctuations) 
are present. 
In this map, we proceed to extract the RS component by 
computing the dipole ($\vec{d}$) and quadrupole ({\boldmath{$Q$}}) moments.
\end{enumerate}
We will adopt here the following definitions for the multipole moments of
a given 2-dimensional map, $M(\vec{x})$, with respect to the coordinate origin: 
\begin{equation}
\vec{d}[M] = \int \frac{\vec{x}}{|\vec{x}|} M(\vec{x}) d^2 x
\end{equation}
\begin{equation}
Q_{ij}[M] = \int \frac{2 x_i x_j - |\vec{x}|^2 \delta_{ij}}{|\vec{x}|^2} 
M(\vec{x}) d^2 x
\label{eq:q_ij}
\end{equation}
Weighting in this way, we 
assure that we give equal weights to equal areas. Thus, this is a
robust approach in the sense that it should tolerate some level of
noise (CMB or instrumental). 

For a 2D image, the {\boldmath{$Q$}}-matrix only has 2 independent
elements because is symmetric, and
tr({\boldmath{$Q$}})=0. Thus, the matrix is specified by quoting 
$Q_{xx}$ and $Q_{xy}$.  However, the quadrupole matrix can be rotated to 
its principal axis, so that a single number characterises the
matrix (one of the eigenvalues), plus an
angle specifying the orientation of the principal axis.  It is easy to
show that the two eigenvalues for {\boldmath{$Q$}} are $\pm
\sqrt{Q_{xx}^2 + Q_{xy}^2}$, which suggests to introduce the total,
coordinate independent quadrupole moment as $ Q = \sqrt{Q_{xx}^2 +
Q_{xy}^2}$. 

We will now illustrate how to extract information about
the physical parameters describing the merger from the suggested
recipe.
For definiteness, we will assume here that we have two observational 
maps available, one of the Compton y-parameter ($y(\vec{x})$), 
and another map of the brightness distribution on the
sky at $\nu=217~GHz$ ($B_{217}(\vec{x})$).
We will denote as ${\bf{Q}^{tSZ}} \equiv {\bf Q}[y(\vec{x})]$ the
quadrupole moment associated to the y-parameter map, and
${\bf{Q}^{T}} \equiv {\bf Q}[B_{217}(\vec{x})]$, the one associated to the 
thermal temperature fluctuation map ($\bf{Q}^{T} =
\bf{Q}^{RS}+ \bf{Q}^{kSZ} + \bf{Q}^{CMB}$). 
We will start by considering the case of zero angular momentum, and 
neglecting the CMB contribution. These other cases will be discussed
in the following two subsections.

\begin{figure}[t]
\centering
\includegraphics[width=\columnwidth]{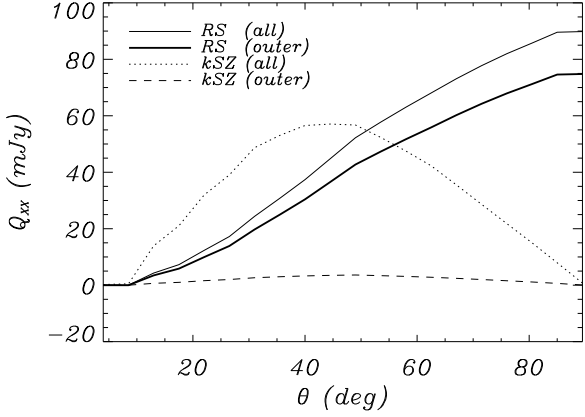}
\caption[]{ Dependence of the measured quadrupole ($Q_{xx}$)
on the projection angle $\ii$. These curves have been derived for
the case of $M_1 = 10^{15} M_\odot$ (left cluster), $M_2 = 5\times
10^{14} M_\odot$ (right cluster), separation of 1~Mpc, $\phi_v =
0\degr$ and the clusters placed along the X-axis 
(so $Q=|Q_{xx}|$) and we use $\ii = \theta_v$ (i.e. zero angular
momentum).  We performed the computation of the quadrupole in the
whole image (quoted as 'all'), and masking the region that in the
y-parameter map has a signal above the 10\% of the peak intensity (quoted as
'outer').  We can see that when doing this, the quadrupole signal is
dominated by the RS component.  
The flux is given in $mJy$, at the 
cross-over frequency of the tSZ effect ($\nu=217~GHz$). The merging
system is assumed to be at $300~Mpc$. 
\label{fig:qxx} }
\end{figure}

In Fig. \ref{fig:qxx} we show the dependence 
of the observed quadrupole with the projection angle. 
The coordinate system has been placed at the 'center-of-light' of
the y-parameter map, and the maps have been rotated so the
cluster centres in the tSZ map lie along the X-axis (so we
only quote $Q_{xx}$ because $Q_{xy}=0$). 
Given that the weighting-factor in Eq. \ref{eq:q_ij} is dimensionless,
then ${\bf{Q}^{T}}$ has units of flux, so
in order to quote these values in $mJy$, we have assumed
that the merging system is located at an angular distance of 300~Mpc.
The map-size used for the computation of the quadrupole was 10.24~Mpc.
This size has been used throughout the paper\footnote{Note that one
expects an small (logarithmic) dependence of the observed quadrupole with the
map size. This can be easily seen for the case of two point-like
clusters, where the asymptotic behaviour of the $\delta T_{RS}$ signal
goes as $r^{-2}$ for large distances $r$ from the center of mass
of the system.}. 
For illustration purposes, the signal coming from the RS and the kSZ 
components has been computed separately, although the observable is
the sum of the two quantities. Given that the RS signal has
a more extended pattern than the kSZ, we have used a mask based on
the tSZ map in order to enhance the signal from the
RS effect. In this example, this mask was obtained by tapering all pixels 
with a temperature in the y-parameter map larger than 10\% of the 
peak temperature.
We can see that when using this taper, the quadrupole becomes
totally dominated by the RS contribution.

\begin{figure}[t]
\centering
\includegraphics[width=\columnwidth]{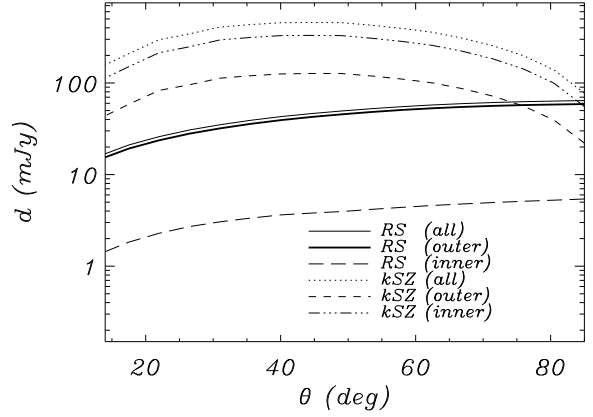}
\caption[]{ Dependence of the measured dipole modulus 
($d=\sqrt{d_x^2+d_y^2}$) with the projection angle $\ii$. 
We consider exactly the same case and notation as in
Fig. \ref{fig:qxx}, but here we also show the dipole
signal from the region where the 90\% of the tSZ emission comes
from (quoted as 'inner'), as well as the complementary 'outer' region. 
\label{fig:d} }
\end{figure}

For illustration purposes, 
we also present in Fig. \ref{fig:d} the same plot but for the dipole
modulus ($d = |\vec{d}| = \sqrt{d_x^2 + d_y^2}$). 
We can see that this dipole signal 
is dominated by the kSZ contribution, and even when we mask 90\% of
the tSZ emission (note that the dipole signal from kSZ is also
reduced by approximately the same factor) the signals from kSZ and
RS are comparable, but in this latter case
only for low values of the projection angle ($\ii \le 50\degr$).
For (close to) face-on mergers, the kSZ contribution goes
down because of the geometrical projection of the velocity 
($v_z \propto \cos \ii$) while the RS component is non zero
($v_\perp \propto \sin \ii$). Therefore, as expected, we conclude that
the quadrupole moment is better suited for the extraction of
the RS signal, while the dipole moment is more suitable for the
kSZ (or tSZ) component (note that when considering only the inner
part of the map, where 90\% of the tSZ emission comes from, then
the dipole is totally dominated by the kSZ component). 

\begin{figure}[t]
\centering
\includegraphics[width=\columnwidth]{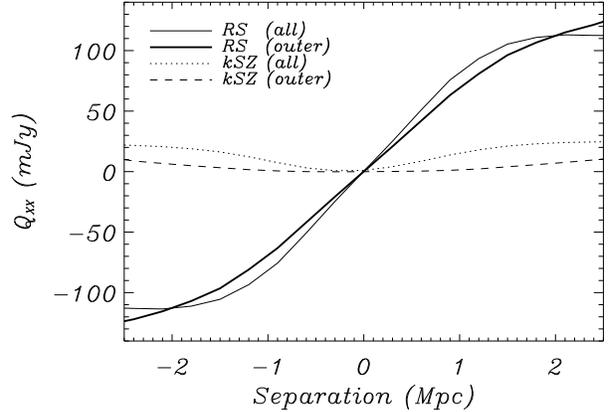}
\caption[]{ Dependence of the measured quadrupole ($Q_{xx}$)
on the separation between clusters. We consider the case 
$M_1 = 10^{15} M_\odot$, 
$M_2 = 5\times 10^{14} M_\odot$, 
$\phi_v = 0\degr$ (so $Q_{yy}=0$) and we use $\ii = \theta_v=80\degr$ 
(zero angular momentum). 
Notation and units are the same as in Fig. \ref{fig:qxx}. Separations are
measured from the most massive to the less massive cluster, 
so in this picture a positive separation represents a 
pre-merger state, where the two clusters are approaching each other, while a
negative separation represents a post-merger state.
\label{fig:qxx2} }
\end{figure}

In Fig. \ref{fig:qxx2} we show the dependence of $Q_{xx}$ on the
physical separation, keeping fixed the other parameters. 
Here the separations are measured from $M_1$ 
(the most massive cluster, i.e. the one with larger tSZ signal) to
$M_2$, and the velocity for $M_1$ is taken to be pointing 
towards the positive direction of X axis.
Within this picture, a positive separation (i.e. cluster $M_1$ on
the left) represents a pre-merger state, where
the two clusters are approaching each other, while a negative
separation (i.e.cluster $M_1$ on
the right) represents a post-merger state.
Thus, we see that the sign of the quadrupole components carries
information about the merger kinematics. 
We also stress here that tapering the central region of the map makes the 
quadrupole dominated by the RS signal. 
Finally, it is worth remarking that depending on the particular
configuration of the merging system, the RS signal adds-up coherently 
or cancels in the computation of the quadrupole (or dipole), 
so we find that the outer region has a larger quadrupole-flux 
than the total map for some physical separations.

\subsection{Characterising the Angular Momentum}

The quadrupole pattern of the RS effect associated to merging clusters
can also be used to learn about the angular momentum of the system, at
least in the plane of the sky: it is clear from 
Figs.~\ref{fig:example1}, \ref{fig:example2} that the direction of
cluster velocities conditions the nature of the quadrupole pattern
imprinted by the RS effect.

This is the motivation of the following approach, which is entirely
based in the computation of the moments of the kSZ, tSZ and RS
components, and hence is well defined for any morphology of the
system:
\begin{itemize}
\item First, one computes the quadrupole moment ${\bf{Q}^{tSZ}}$ 
(it would be valid to  compute the direction associated to the 
dipole of the kSZ map, however, we shall ignore it due to its 
dependence on the inclination angle). 
From this moment, one computes the following angle, corresponding to one
of the principal directions of the quadrupole: 
\begin{equation}
\alpha_{\bf{Q}^{tSZ}}  =  \arctan{ \left(\frac{Q^{tSZ}_{xy}}
 {Q^{tSZ}_{xx}-\sqrt{\left(Q^{tSZ}_{xy} \right)^2 + 
	\left(Q^{tSZ}_{xx} \right)^2 }}\right) }. 
\label{eq:angles1}
\end{equation}
\item Second, one computes the quadrupole moment ${\bf{Q}^{T}}$, 
which should be dominated by the RS effect due to the tapering of the central
  region. It is then convenient to take the angle related
  to the positive eigenvalue:
\begin{equation}
\alpha_{\bf{Q}^{T}}  =  \arctan{\left( \frac{Q^{T}_{xy}}
 {Q^{T}_{xx}+\sqrt{\left(Q^{T}_{xy} \right)^2 + 
	\left(Q^{T}_{xx} \right)^2 }}\right) }, 
\label{eq:angles2}
\end{equation}
\end{itemize}
Taking as the origin of our coordinate frame the centre of {\it
tSZ-light} of the system, the first angle should be related to the 
angle of the direction joining the centres of mass of each cluster
($\alpha_{CM}$). Indeed, we find that 
$\alpha_{\bf{Q}^{tSZ}} \simeq \alpha_{CM}$.  
In a similar way, the angle $\alpha_{\bf{Q}^{T}}$ is
related to the direction of the individual velocities of the clusters
via $\alpha_{\bf{Q}^{T}} \simeq \phi_{v}/2$.
Thus, we can use the two observables given in Eq. \ref{eq:angles1}
and \ref{eq:angles2} to directly estimate the two physical angles
$\alpha_{CM}$ and $\phi_v$. 

\begin{figure}[t]
\includegraphics[width=\columnwidth]{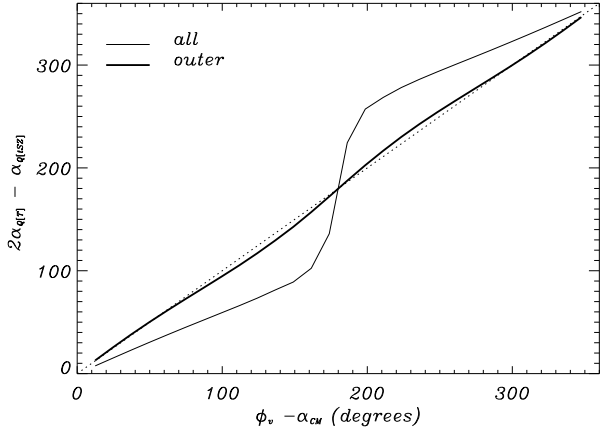}
\caption[]{  Recovery of $\phi_{v} - \alpha_{CM}$ from
$2 \alpha_{\bf{Q}^{T}} - \alpha_{\bf{Q}^{tSZ}} $.
Thick solid line excludes that area containing 90\% of the tSZ emission,
whereas thin solid line considers the entire map. The dotted line gives
the expected scaling $2 \alpha_{\bf{Q}^{T}} - \alpha_{\bf{Q}^{tSZ}} = 
\phi_v - \alpha_{CM}$. A projection
angle of $\ii = 60\degr$ was used in the calculation. }
\label{fig:angles} 
\end{figure}

\begin{figure}[t]
\includegraphics[width=\columnwidth]{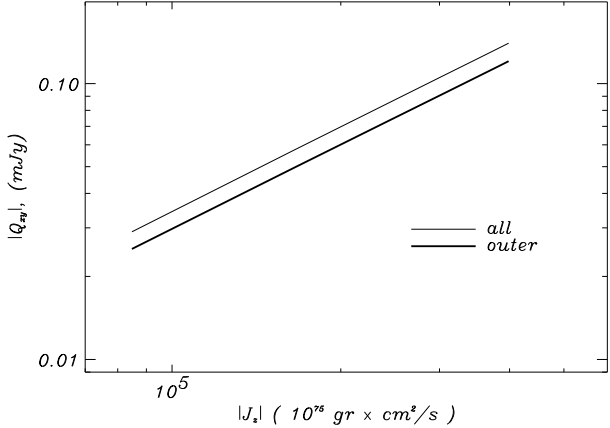}
\caption[]{  Dependence of
$|Q^{T}_{xy} |$ with the absolute value of the 
projected component of the angular
momentum, $|J_z|$. We use the same physical parameters of the
previous figure, and computed for each value of $\phi_v$ 
the projected Z-component of the angular momentum 
(note that, for $\phi_v=0\degr$ and $\phi_v=180\degr$, ${\bf J} = 0$).
Under the symmetry of our system (clusters along the X-axis), the kSZ residual
affects mostly the $Q^{T}_{xx}$ values, leaving the crossed term 
$Q^{T}_{xy}$ practically unaltered. } 
\label{fig:J} 
\end{figure}

This is illustrated in Fig. \ref{fig:angles}, where we use the 
cluster parameters \{ $M_1=10^{15}M_{\odot}$, 
$M_2=5\times10^{14}M_{\odot}$, $\ii =\theta_v=60\degr$, $d=1~Mpc$ \}, 
and we vary the azimuthal angle
$\phi_v$ from 0 to $2\pi$ radians. The quadrupole induced by the kSZ 
component perturbs our estimate of $\phi_v$ from $2
\alpha_{\bf{Q}^{T}}$ as it is shown in the figure. 
Note that the value of $\phi_v$ at which we have the maximum departure
from the real value is not constant, but depends on the relative 
amplitude ratio of the kSZ and RS quadrupoles,
i.e. depends on the other physical parameters of the merger. 
Finally, we again obtain better results for 
the case in which we mask out the area where 90\% of the tSZ emission
comes from, 
even for cases with low values of the projection angle.

With respect to the amplitude of the angular momentum, one can see
that it is closely related to the cross component of the $\bf{Q}^{RS}$
tensor, provided the X-axis of our coordinate system is located along
the axis joining the cluster centres, and its centre is the system
centre of mass. In this reference frame, a merger with $\bf{J}=0$
shows no $Q^{RS}_{xy}$ component.  Indeed, for the case of two
point-like clusters, it can be shown that $Q^{RS}_{xy} = J_z\;
f(x_1,x_2)$, with $x_1$,$x_2$ the coordinates of the cluster centres, 
and $J_z$ the projected component of the angular momentum along the
Z-axis (see Appendix). Therefore, one can make use of the transformations
of the components of the quadrupole under a 
rotation of the coordinate system by an angle $\alpha$,
\[
{\tilde Q}_{xx} = \cos (2\alpha) \; Q_{xx} - \sin (2\alpha) \; Q_{xy}
\]
\begin{equation}
{\tilde Q}_{xy} = \sin (2\alpha) \; Q_{xx} + \cos (2\alpha) \; Q_{xy}.
\label{eq:transQ}
\end{equation}
in order to relate $J_z$ with the observable quantities 
$(Q^{T}_{xx},Q^{T}_{xy})$ computed in an arbitrarily oriented
reference frame.  Using the observable $\alpha_{{\bf Q^{tSZ}}}$ to
estimate the orientation of the axis joining the
cluster mass centres ($\alpha_{CM}$), then 
the quadrupole components in the reference system
aligned with the mass centre axis can be computed from
Eq. \ref{eq:transQ} using $\alpha=\alpha_{{\bf Q^{tSZ}}}$. 
As we have seen above, for high values
of the projection angle $\ii$, we have that ${\tilde Q}^{T}_{xy} \approx 
{\tilde Q}^{RS}_{xy}$, and the latter is 
proportional to the magnitude of $J_z$. 
In Fig. \ref{fig:J} we illustrate 
this dependence of ${\tilde Q}^{RS}_{xy}$
versus the Z-projection of the angular momentum, for the
same physical parameters as in Fig. \ref{fig:angles}.

\subsection{Stacking of clusters}

The RS effect produces only a small temperature fluctuation making it
very hard to measure it at individual clusters. It is therefore worth
to develop ideas on how to co-add the signals from a large sample of
clusters leading to a statistical RS effect detection. 

The simplest approach could be to co-add the quadrupole strength $Q$,
which is always positive. However, there are two problems with
this. First, the intrinsic CMB temperature fluctuations 
contribute also to $Q$, although their contribution can relatively accurately
be estimated by measuring at non-cluster positions. Second, the kSZ
effect imprints some quadrupole structure, which can not be completely
masked away. Estimates of its strength are not possible from the
observational data alone and therefore would rely on numerical
simulations of galaxy cluster merger in a cosmological setting.

We therefore propose a different approach to stack the signal from
several clusters. First, the cluster should be rotated in a way that
its major axis (of the tSZ effect or X-ray image) is aligned with the
X-axis. Then, the quadrupole tensor of the temperature fluctuations
${\bf Q^{T}} = {\bf Q^{CMB}} + {\bf Q^{RS}} + {\bf Q^{kSZ}}$ is
calculated in this coordinate system and its components are co-added
for all the clusters in the sample. We expect that for a sufficiently
large number of clusters all contributions to ${\bf Q^{T}}$ cancel out
due to symmetry properties except the ones from the RS effect. 

The components of ${\bf Q^{CMB}}$ have both signs with equal
probability, since the CMB fluctuations are not correlated with the
foreground cluster and the signs of quadrupole components are not
rotational invariant (they reverse after a $90^\circ$ rotation).
For the same reason, we expect that the lensing-induced cluster signal
in the CMB \citep{2000ApJ...538...57S} will also cancel out, because
the quadrupole moment is a linear function of the temperature, for 
which we expect both signs with equal probability. 

Also the components of ${\bf Q^{kSZ}}$ have both signs with equal
probability: mirroring a cluster through the X-Y plane going through
its centre of mass reverses the sign of the kSZ map, and due to
linearity also the sign of all ${\bf Q^{kSZ}}$ components. Since a
cluster and its mirrored counterpart are equally likely in the cluster
sample, their contributions to the co-added quadrupole cancel each other
statistically.

The components of ${\bf Q^{RS}}$ have both signs with different
probabilities. Since galaxy clusters are growing, one should find more
clusters with converging velocity fields than with diverging
ones. Diverging velocity fields should be present shortly after core
passage of the dark matter clumps of two merging clusters. However,
dynamical effects due to gravitational interactions (violent
relaxation) should convert rapidly the organised bulk motion of the
dark matter particles into an undirectional motion.
This reduces the statistical contribution of diverging matter flows to the
co-added ${\bf Q^{T}}$ compare to the contribution of the converging
matter flows of early stage mergers.

For this reasons, only a contribution from the RS effect due to the
growing large-scale structure should survive statistically. However,
there is one possible contamination to the described measurement
scheme: any residual tSZ contamination to the temperature
fluctuations is expected to be co-aligned with the major axis, and
therefore to add coherently to ${\bf Q^{T}}$, unless they appear with
both signs with equal probability. This point could be achieved by
taking relativistic corrections to the tSZ effect into account.

Therefore, further theoretical investigations are required before this
stacking method can be fully be established. This issue will be
addressed in a follow-up paper.

\section{Conclusions}

In this paper we have presented a formalism that can be easily
incorporated to N-body simulation codes in order to predict
the Rees-Sciama effect in a merging systems of clusters of galaxies. 
For the typical range of velocities in those systems ($v/c \ll 1$), the
obtained expression (Eq. \ref{eq:dTRS2}) can also be seen 
as a gravitational lensing effect produced by a moving lens.

Using simple modelling for the cluster merger, we have illustrated
the morphology and symmetries of this effect, and we have developed a
method to extract the signal, which should be applicable to realistic
maps. This method is based on the computation of the (weighted) dipole
and quadrupole moments of the brightness distribution. In particular,
it has been shown how the quadrupole moment is related to the
kinematic properties of the merger, so we can extract information
about the dynamical state of the system: pre-merger or 
post-merger, and the magnitude of the angular momentum.

Since we expect that in the near future the observation of a single
cluster merger will be extremely difficult, given the weak signal
strength, we have proposed a simple method of stacking the signal from a
large number of clusters in order to extract their RS signature 
statistically. The procedure is straight-forward: for a sample of
clusters (e.g. a complete, or a merging cluster sample) a coordinate
system is attached to the center of light (X-ray or tSZ map) so that
the X-axis is aligned with the major elongation of the gas. Then the
quadrupole moments of the CMB temperature fluctuations (outside the
gas region, defined by the tSZ effect) are calculated for each
cluster. Finally the individual quadrupole moment components are
co-added for the sample. Intrinsic CMB fluctuations and kSZ effect
residuals should cancel out statistically, leaving only a signature of
the RS effect.  A detailed investigation of this
stacking method will be carried out in a follow-up paper.

\begin{acknowledgements}
We acknowledge useful comments by S.~D.~M. White and M. Bartelmann. 
JARM and CHM acknowledge the financial support provided through the European
Community's Human Potential Programme under contract HPRN-CT-2002-00124, 
CMBNET. We thank an anonymous referee for useful comments.
\end{acknowledgements}


\appendix

\section{Relationship between the quadrupole moment and the angular momentum}
In this section, we prove that for two point-like clusters,
$Q^{RS}_{xy}$ is directly proportional to the component
of the angular momentum projected along the line of sight, provided
the coordinate origin is located at the mass centre
of the system and X-axis is aligned to the axis joining the clusters.
Thus, the z=0 plane is the image plane. 

Let us assume that the coordinates of clusters are given by 
$\vx_i\equiv (x_i,0)$, for $i=1,2$. Their linear momenta will be labelled as 
${\bf P}_i \equiv M_i {\bf v}_i$, $i=1,2$ and we set the total
momentum of the system to zero, so 
${\bf P}_1=-{\bf P}_2$. Also note that from the choice of the
coordinate system, then $\vx_1$ and $\vx_2$ are related.
The {\it xy} component of the RS-induced quadrupole is then given by:
\begin{equation}
Q^{RS}_{xy}= \frac{-4G}{c^3}\int d^2 x\; \int  d^2 x' \;\frac{2 x y}{\vx^2}\;
  {\tilde{\bf p}}_\perp(\vxp) \vec{\cdot} \frac{\vx-\vxp}{(\vx-\vxp)^2},
\label{eq:demQxy1}
\end{equation}
with ${\tilde{\bf p}}_\perp(\vxp)$ the projection on the z=0 plane of
the LOS projected momentum density, given in
Eq.~\ref{eq:dTRS2}. For our two point masses, we can take 
${\tilde{\bf p}}_\perp (\vxp) = 
{\bf P}_1 \delta_D (\vxp - \vx_1) + {\bf P}_2 \delta_D
(\vxp - \vx_2) $, $\delta_D$ denoting Dirac delta. It yields
\begin{equation}
Q^{RS}_{xy}= \frac{-8G}{c^3}\int d^2 x \;\frac{x y}{\vx^2}\; {\bf P}_1
\left[ \frac{\vx-\vx_1}{|\vx-\vx_1|^2} -\frac{\vx-\vx_2}{|\vx-\vx_2|^2}
			\right]. 
\label{eq:demQxy2}
\end{equation}
The scalar product in this equation can be decomposed in two terms:
\[
Q^{RS}_{xy}= \frac{-8G}{c^3}\int d^2 x \; \frac{x y}{\vx^2}\biggl\{
	 P_{1,x}
			 \biggl( \frac{x-x_1}{(x-x_1)^2+y^2} -
\]
\[
\phantom{xxxxxxxxxxxxxx}
\frac{x-x_2}{(x-x_2)^2+y^2} \biggr) \; +
\]
\[
\phantom{xxxxxxxx} 
 P_{1,y}	\biggl( \frac{y}{(x-x_1)^2+y^2}\;-
\]
\begin{equation}
\phantom{xxxxxxxxxxxxxx} 
\frac{y}{(x-x_2)^2+y^2}	\biggr) \biggr\}.
\label{eq:demQxy3}
\end{equation}
The term proportional to $P_{1,x}$ vanishes when integrating over $y$, 
because the integrand is an odd function on $y$. 
The remaining term can easily
be rewritten as:
\[
Q^{RS}_{xy}= J_z \; \frac{(-8G)}{c^3}\int d^2 x \;\frac{x\;y^2}{\vx^2}
			\; \times
\]
\begin{equation}
\phantom{xxxxxxxxxx}
 \biggl\{
	\frac{(x_1-x+x_2-x)}{\left[(x-x_1)^2+y^2\right]
							\left[ (x-x_2)^2+y^2\right]} \biggr\},
\label{eq:demQxy4}
\end{equation}
with $J_z \equiv  -P_{1,y} (x_2 - x_1)$ the LOS projected component
of the angular momentum. Therefore,
\begin{equation}
Q^{RS}_{xy}= J_z\; f(x_1,x_2),
\label{eq:demQxy4}
\end{equation}
with $f(x_1,x_2)$ a function depending exclusively on the geometry 
of the system.


\end{document}